\documentclass[aps,prb,reprint,showpacs]{revtex4-1}
\usepackage{amsmath,amssymb,mathrsfs}
\usepackage[dvips]{graphicx}

\newcommand{\ii}{\text{i}}

\allowdisplaybreaks[4]

\begin{document}
\title{Spectral properties of one-dimensional spiral spin density wave states}
\author{Dirk Schuricht}
\affiliation{Institute for Theory of Statistical Physics, 
RWTH Aachen, 52056 Aachen, Germany}
\affiliation{JARA-Fundamentals of Future Information Technology}
\date{\today}
\pagestyle{plain}

\begin{abstract}
We provide a full characterization of the spectral properties of spiral spin density wave (SSDW)
states which emerge in one-dimensional electron systems coupled to localized magnetic 
moments or quantum wires with spin-orbit interactions. 
We derive analytic results for the spectral function, local density of states and 
optical conductivity in the low-energy limit by using field theory techniques. We identify 
various collective modes and show that the spectrum strongly depends on the interaction 
strength between the electrons. The results provide characteristic signatures for an 
experimental detection of SSDW states.
\end{abstract}
\maketitle

It is by now well established that interactions have drastic effects on one-dimensional 
systems.\cite{Giamarchi04} For example, the spectral properties of Luttinger liquids, 
which one may regards as the standard model of one-dimensional electrons,
show spin-charge separation as well as non-trivial power laws which depend on 
the interaction strength.\cite{LL} The system
decomposes into two gapless sectors describing collective charge and spin excitations
which possess linear dispersion relations with velocities $v_\text{c}$ and $v_\text{s}$,
respectively. More general situations can be described by adding appropriate terms to 
the Luttinger liquid Hamiltonian. For example, spin-orbit interactions (SOI) lead to a mixing 
of charge and spin degrees of freedom, which strongly affects the spectral 
function but leaves the system gapless.\cite{Schulz-09}

A more profound change in the low-energy properties happens when the system develops
a gap. The prototype scenario\cite{Giamarchi04} for this situation is the Mott transition 
at which a gap in the charge sector opens provided the system is at commensurate filling. 
The spectral function of the resulting Mott insulator (MI)\cite{MI} 
still shows spin-charge separation but the charge excitations possess a mass $\Delta$. 
A similar situation is encountered in systems with SOI and an additional 
magnetic field, where a spin gap opens while the charge sector remains 
gapless.\cite{Sun-07} On the other hand, if the SOI
are spatially modulated with a wave length commensurate with the Fermi wave length,  
the system develops gaps in both sectors.\cite{Japaridze-09}
In all three situations the system separates into charge and spin sectors, i.e., 
there is no non-trivial mixing of gapless and gapped degrees of freedom.

As it was shown by Braunecker \emph{et al.},\cite{Braunecker-09} this is no longer the 
case if one considers the interplay of one-dimensional electronic degrees of freedom
and localized magnetic moments such as nuclear spins. In such systems feedback effects 
between the conduction electrons and nuclear magnetic moments can, at sufficiently low 
temperatures, stabilize the coexistence of electron density order and helical magnetic order. 
Specifically they showed that the electronic subsystem develops a gap in a superposition
of charge and spin degrees of freedom accompanied by a spiral spin density 
wave (SSDW)\cite{SLL}  
following the helical magnetic order. As the original electrons possess a highly 
non-trivial decomposition into the gapped and gapless sectors one expects the system 
to possess rather unusual spectral properties. We note that similar situations are  
encountered in Luttinger liquids coupled to extended ferromagnetic domain 
walls\cite{PereiraMiranda04} as well as in quantum wires with SOI in a magnetic 
field along the wire provided the band shift induced by the SOI is commensurate 
with the Fermi momentum.\cite{Braunecker-10}

In this Rapid Communication we provide a detailed analysis of the spectral properties 
of such SSDW states. We show that the spectral function is strongly asymmetric 
in the spin components and identify various dispersing collective 
modes. The calculated local density of states (LDOS) exhibits typical features of both
the Luttinger liquid and the MI. In particular we are able to derive 
exact results for the appearing power-law exponents. Finally we show that the optical 
conductivity possesses characteristic singularities intimately related to the interactions
between the electrons. Taken together our results provide a set of specific signatures 
necessary for the experimental identification of SSDW states.

\emph{Model.} Experimental systems in which such states may exist require the 
interaction between electrons and nuclear spins in one dimension as it is, for example,
the case in $^{13}$C single-wall carbon nanotubes\cite{CNT} or GaAs quantum 
wires.\cite{GaAs}
Braunecker \emph{et al.}\cite{Braunecker-09} showed that in these systems
the conduction electrons mediate 
an RKKY 
interaction between the nuclear spins.
The feedback of the resulting helical magnetic order causes the formation of a SSDW
state and generates a gap in the electronic subsystem, which can be 
described by the effective low-energy Hamiltonian
(which is also relevant\cite{Braunecker-10} to quantum wires with SOI showing 
a spin-selective Peierls transition)
\begin{eqnarray}
H&=&H_++H_-,\label{eq:hamiltonian}\\
H_+&=&\frac{v_+}{16\pi}\int \text{d}x 
\Bigl[\bigr(\partial_x\Phi_+\bigr)^2+
\bigr(\partial_x\Theta_+\bigr)^2\Bigr]\nonumber\\*
&&+\frac{B}{2\pi}\int \text{d}x\,
\cos\bigl(\sqrt{K/4}\,\Phi_+\bigr),
\label{eq:plushamiltonian}\\
H_-&=&\frac{v_-}{16\pi}\int\text{d}x 
\Bigl[\bigr(\partial_x\Phi_-\bigr)^2+
\bigr(\partial_x\Theta_-\bigr)^2\Bigr].
\label{eq:minushamiltonian}
\end{eqnarray}
Here $\Phi_\pm$ are canonical Bose fields and $\Theta_\pm$ are their dual fields.
The velocities $v_\pm$ and the Luttinger parameter $K$ are related to the original 
parameters in the charge and spin sectors via 
\begin{equation}
v_+=\frac{1}{K}\left[v_\mathrm{c}K_\mathrm{c}+\frac{v_\mathrm{s}}{K_\mathrm{s}}\right],\quad
v_-=\frac{1}{K}\left[\frac{v_\mathrm{c}}{K_\mathrm{s}}+v_\mathrm{s}K_\mathrm{c}\right],
\end{equation}
and $K=K_\text{c}+1/K_\text{s}$. For $K<4$ the cosine term in \eqref{eq:plushamiltonian}
is relevant and generates a gap $\Delta$, which is a function\cite{Zamolodchikov95} of 
the bare Overhauser field $B$ and $K$. We will view it here as a
phenomenological parameter replacing $B$ such that we are left with the
four free parameters $K_\text{c,s}$, $v_\text{c}/v_\text{s}$, and $\Delta$. For repulsive 
electron-electron interactions one has $0<K_\text{c}<1$. Furthermore, in the 
derivation of \eqref{eq:hamiltonian} it was assumed 
that\cite{Braunecker-09} $K_\text{s}\ge 1$, 
where $K_\text{s}=1$ corresponds to the spin rotational invariant situation.

The bosonic fields appearing in \eqref{eq:hamiltonian} are related to the slowly varying 
right- and left-moving Fermi fields defined via 
\begin{equation}
\label{eq:lowenergy}
\Psi_{\sigma}(x)=e^{\ii k_\text{F}x}\psi_{\text{R}\sigma}(x)
+e^{-\ii k_\text{F}x}\psi_{\text{L}\sigma}(x),
\end{equation}
by the relations\cite{Braunecker-09} 
($l=\text{R/L}\equiv\pm$, $\sigma=\uparrow,\downarrow\equiv\pm$)
\begin{eqnarray}
\label{eq:bosonization1}
\psi_{\pm\pm}&=&\frac{\eta_{\pm\pm}}{\sqrt{2\pi}}\,
\exp\bigl(\mp\ii a\Phi_+\bigr)\,\exp\bigl(\pm \ii b\Phi_--\ii c\Theta_-\bigr),\\
\psi_{\pm\mp}&=&\frac{\eta_{\pm\mp}}{\sqrt{2\pi}}\,
\exp\bigl(\tfrac{\ii}{2\sqrt{K}}\Theta_+\mp\ii\tfrac{\sqrt{K}}{4}\Phi_+\bigr)\,
\exp\bigl(\ii d\Theta_-\bigr),\quad\label{eq:bosonization2}
\end{eqnarray}
where the Klein factors $\eta_{l\sigma}$ satisfy the anticommutation relations
$\{\eta_{l\sigma},\eta_{l'\sigma'}\}=2\delta_{ll}\delta_{\sigma\sigma'}$ 
and the constants are defined by $a=(K_\text{c}K_\text{s}-1)/(4K_\text{s}\sqrt{K})$, 
$b=\sqrt{K_\text{c}/(KK_\text{s})}/2$, and 
$c,d=(K_\text{c}K_\text{s}\pm 1)/(4\sqrt{KK_\text{c}K_\text{s}})$.

The excitations in the sine-Gordon model \eqref{eq:plushamiltonian} are solitons 
and antisolitons with mass $\Delta$, and, if $K<4/(N+1)$, breather (soliton-antisoliton)
bound states $B_n$, $n=1,\ldots,N$, with masses $\Delta_n=2\Delta\sin(n\pi K/(8-2K))$. 
We note that there exists at least one breather for repulsive interactions. The solitons 
and antisolitons possess a relativistic dispersion relation $E^2=\Delta^2+v_+^2P^2$;
the dispersion relation for the $n$th breather is obtained by replacing $\Delta$ by
$\Delta_n$. The sector \eqref{eq:minushamiltonian} describes gapless collective 
excitations propagating with velocity $v_-$.

\emph{Green's function.} The spin-resolved spectral function and LDOS discussed below 
follow from the spin-diagonal Green's function in Euclidean space, which is defined by
\begin{equation}
\label{eq:GF}
G_{\sigma}(\tau,x)=\sum_{ll'}e^{\ii lk_\text{F}x}\,G_{\sigma}^{ll'}(\tau,x),
\end{equation}
where $G^{ll'}_\sigma(\tau,x)= -\langle\mathcal{T}_\tau\,\psi_{l\sigma}(\tau,x)\,
\psi_{l'\sigma}^\dagger(0,0)\rangle$, 
$\tau=\ii t$ denotes Euclidean time, and $\mathcal{T}_\tau$ is the $\tau$-ordering operator.
Due to the decomposition of the Hamiltonian the Green's 
function factorizes into a product of correlation functions in the gapped and gapless
sectors. The correlation functions in the gapless sector can be straightforwardly 
calculated by using standard methods.\cite{Giamarchi04} The derivation in the gapped 
sector employs the integrability of the sine-Gordon model 
via a form-factor expansion\cite{Smirnov92book,EsslerKonik05} with the results (${\tau>0}$)
\begin{equation}
  \begin{split}
  &G^\text{RR}_{\uparrow}(\tau,x)=-\frac{1}{2\pi}
  \frac{1}{(v_-\tau-\ii x)^{2(b+c)^2}}\frac{1}{(v_-\tau+\ii x)^{2(b-c)^2}}\\
  &\;\times G_a^2\left[1+
  \frac{\lambda^2\sin^2\tfrac{2\pi\sqrt{K}a}{4-K}}{\pi\sin^2\tfrac{\pi K}{4-K}}
   K_0\Bigl(\Delta_1\sqrt{\tau^2+x^2/v_+^2}\Bigr)+\ldots\right]
  \end{split}
  \label{eq:G1}
\end{equation}
and
\begin{equation}
  \begin{split}
  &G^\text{RR}_{\downarrow}(\tau,x)=-\frac{1}{2\pi}
  \frac{1}{(v_-\tau-\ii x)^{2d^2}}\frac{1}{(v_-\tau+\ii x)^{2d^2}}\\
  &\;\times\left[\frac{Z_1}{\pi}\left(\frac{v_+\tau+\ii x}{v_+\tau-\ii x}\right)^{\tfrac{1}{4}}
  K_1\Bigl(\Delta\sqrt{\tau^2+x^2/v_+^2}\Bigr)
  +\ldots\right].
  \end{split}
  \label{eq:G2}
\end{equation}
Here $K_{0/1}$ denote modified Bessel functions\cite{AS} and the constants $G_a$, 
$\lambda$, and $Z_1$ (which depend on $K_\text{c}$ and $K_\text{s}$) were obtained in 
Refs.~\onlinecite{LukyanovZamolodchikov97,FF}.
The subleading terms in \eqref{eq:G1} and \eqref{eq:G2} involve heavier breathers $B_n$,
$n\ge 2$ (for $K<4/3$), or multi-particle states. Furthermore we have 
$G_{\sigma}^\text{LL}(\tau,x)=G_{\bar{\sigma}}^\text{RR}(\tau,-x)$, 
$G_{\sigma}^{ll}(-\tau,-x)=G_{\sigma}^{ll}(\tau,x)$, and
$G_{\sigma}^\text{RL}=G_{\sigma}^\text{LR}=0$ with
$\bar{\sigma}=\mp$ for $\sigma=\pm$.

\emph{Spectral function.} The spectral function can be experimentally probed
using angle-resolved photoemission measurements. We can obtain the spin-resolved 
spectral function $A_\sigma(\omega,p)=A^>_\sigma(\omega,p)+A^<_\sigma(\omega,p)$
directly from the Green's function \eqref{eq:GF} via
\begin{equation}
A^\gtrless_\sigma(\omega,p)=\mp\int\frac{\text{d}t}{2\pi}\frac{\text{d}x}{2\pi}\,
e^{\ii(\omega t-px)}\,G_\sigma(\tau\gtrless 0,x)\Bigl|_{\tau\to\ii t\pm\delta},
\label{eq:SF}
\end{equation}
where $\delta$ is small but finite and mimics broadening by instrumental resolution 
and temperature in experiments. The form-factor expansions \eqref{eq:G1} and 
\eqref{eq:G2} result in a similar expansion 
for the spectral function, where each term can be expressed as an integral over a
hypergeometric function depending on the parameters $b,c$, and $d$. 
We note that the multi-particle corrections to \eqref{eq:G1} and \eqref{eq:G2} yield 
negligible\cite{FFexpansion} contributions 
at low energies and vanish for $|\omega|<2\Delta$. The spectral
function satisfies $A_\sigma(-\omega,k_\text{F}-q)=A_\sigma(\omega,k_\text{F}+q)$
and $A_\sigma(\omega,-k_\text{F}-q)=A_{\bar{\sigma}}(\omega,k_\text{F}+q)$.

\begin{figure}[tb]
\centering
\includegraphics[scale=0.35,clip=true]{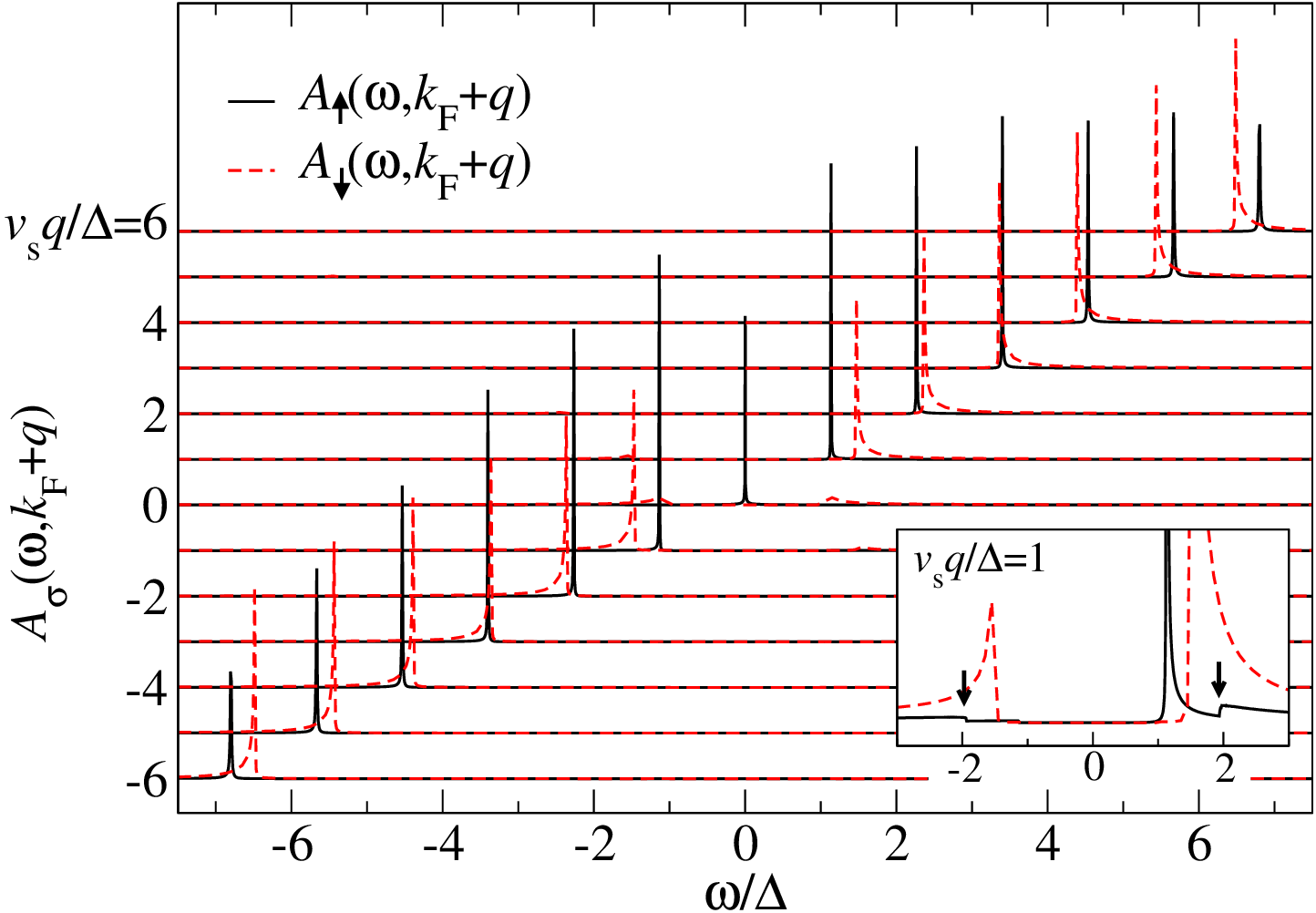}
\caption{(Color online)
Spectral function $A_\sigma(\omega,k_\text{F}+q)$ for $K_\text{c}=0.5$, $K_\text{s}=1$, 
$v_\text{c}=1.2\,v_\text{s}$, 
$\sigma=\uparrow$ (solid black lines), and $\sigma=\downarrow$ (dashed red lines). 
The curves are constant $q$-scans which have been offset along the $y$-axis with 
respect to one another. We stress that the up-spin 
component possesses spectral weight within the gap $\Delta$ and that the dispersions
cross at $v_\text{s}q_0/\Delta\approx\pm 2.6$. The arrows in the inset
indicate the energies where processes involving $B_1$ set in.}
\label{fig:A}
\end{figure}
The spectral function in the vicinity of $p=k_\text{F}$ is shown in Fig.~\ref{fig:A}.
We observe a strong spin dependence. The up-spin component is dominated by a 
linearly dispersing feature following $\omega=v_-q$ which originates from the gapless 
excitations of \eqref{eq:minushamiltonian}. In addition the existence of 
breathers results in dispersing features at $\omega=\pm(\Delta_n^2+v_+^2q^2)^{1/2}$ 
which are, however, very weak (see the arrows in the inset in Fig.~\ref{fig:A}). 
From our analytical results we can extract the power laws at the dispersing features,
e.g., for fixed $q>0$ and $\delta\to 0^+$, we obtain
\begin{equation}
A_\uparrow(\omega,k_\text{F}\pm q)\propto(\omega-v_-q)^{2(b\mp c)^2-1},
\;\omega\to v_-q^+,
\label{eq:pl1}
\end{equation}
while at $q=0$ the exponent is $4(b^2+c^2)-2$. 
The measurement of these power laws can be used to extract the interaction strength
encoded in the Luttinger parameters $K_\text{c}$ and $K_\text{s}$.
We note that the derivation\cite{Braunecker-09} of the effective low-energy model 
\eqref{eq:hamiltonian} relied on a linear single-particle dispersion, whereas in any realistic
microscopic model some curvature will be present. Although this curvature is irrelevant
in the renormalization-group sense, it has been shown\cite{curvature} to affect the power
laws in the spectral function. However, as was argued in Ref.~\onlinecite{Braunecker-09}
for both single-wall carbon nanotubes and GaAs quantum wires the 
curvature-induced deviations from \eqref{eq:hamiltonian} are negligible. 
We further note that a potentially present momentum dependence of the two-particle 
interaction\cite{Meden99} was neglected in the derivation\cite{Braunecker-09} 
of \eqref{eq:hamiltonian}.

The down-spin component of the spectral function is dominated for $q>0$ by a dispersing 
mode following $\omega=\omega_q\equiv(\Delta^2+v_+^2q^2)^{1/2}$ for which we find
\begin{equation}
A_\downarrow(\omega,k_\text{F}+q)\propto(\omega-\omega_q)^{4d^2-1},
\;\omega\to\omega_q^+.
\end{equation}
In addition, there is a very weak feature at $\omega=-\omega_q$. The dispersing modes 
for $q<0$ are obtained using the symmetry of the spectral function. These dispersing 
features originate from the massive soliton excitations of \eqref{eq:plushamiltonian}, in particular the spectral function vanishes for $|\omega|<\Delta$. 
For generic system parameters one has 
$v_\text{c}>v_\text{s}$ and $K_\text{c}K_\text{s}<1$, which results in a crossing of the 
dispersing up- and down-spin features at $q_0=\pm\Delta/(v_-^2-v_+^2)^{1/2}$. 
In the non-interacting case, ${K_\text{c}=K_\text{s}=1, v_\text{c}=v_\text{s}\equiv v}$, the 
contributions from the breathers vanish and the remaining dispersing features become
$\delta$-functions, e.g., 
${A_\uparrow(\omega,k_\text{F}+q)}\propto\delta(\omega-vq)$.

The vanishing of $A_\downarrow(\omega,k_\text{F}+q)$ and  
$A_\uparrow(\omega,-k_\text{F}-q)$ for $|\omega|<\Delta$ shows that the system 
effectively acts as a spin filter at sufficiently low energies. As was pointed out in 
Ref.~\onlinecite{Braunecker-09}, the existence of a gap in one half of the conduction 
channels yields a reduction of the conductance by a factor of 2. We note, however, 
that an experimental detection in transport experiments may be hampered by the 
influence of the attached contacts.\cite{conductance}

Finally let us compare our results to the one-dimensional 
MI, whose spectral function\cite{MI}
does not depend on the spin, possesses a gap, and clearly displays 
propagating charge and spin excitations. 
Thus although the low-energy theories of the MI and the SSDW state are both 
given by a sum of a gapped and a gapless sector, their spectral functions differ 
significantly due to the non-trivial decomposition of the electronic degrees of 
freedom into elementary excitations in the latter. In particular, the dispersing features in 
the SSDW state do not permit a simple interpretation in terms of individually propagating 
charge and spin excitations, i.e., a direct observation of spin-charge separation in 
a SSDW state is not possible. 
Furthermore the operator $e^{\mp\ii a\Phi_+}$  appearing in  \eqref{eq:bosonization1} 
possesses a non-vanishing vacuum expectation value\cite{LukyanovZamolodchikov97} 
and thus gives rise to spectral weight at energies $|\omega|<\Delta$.

\begin{figure}[tb]
\centering
\includegraphics[scale=0.35,clip=true]{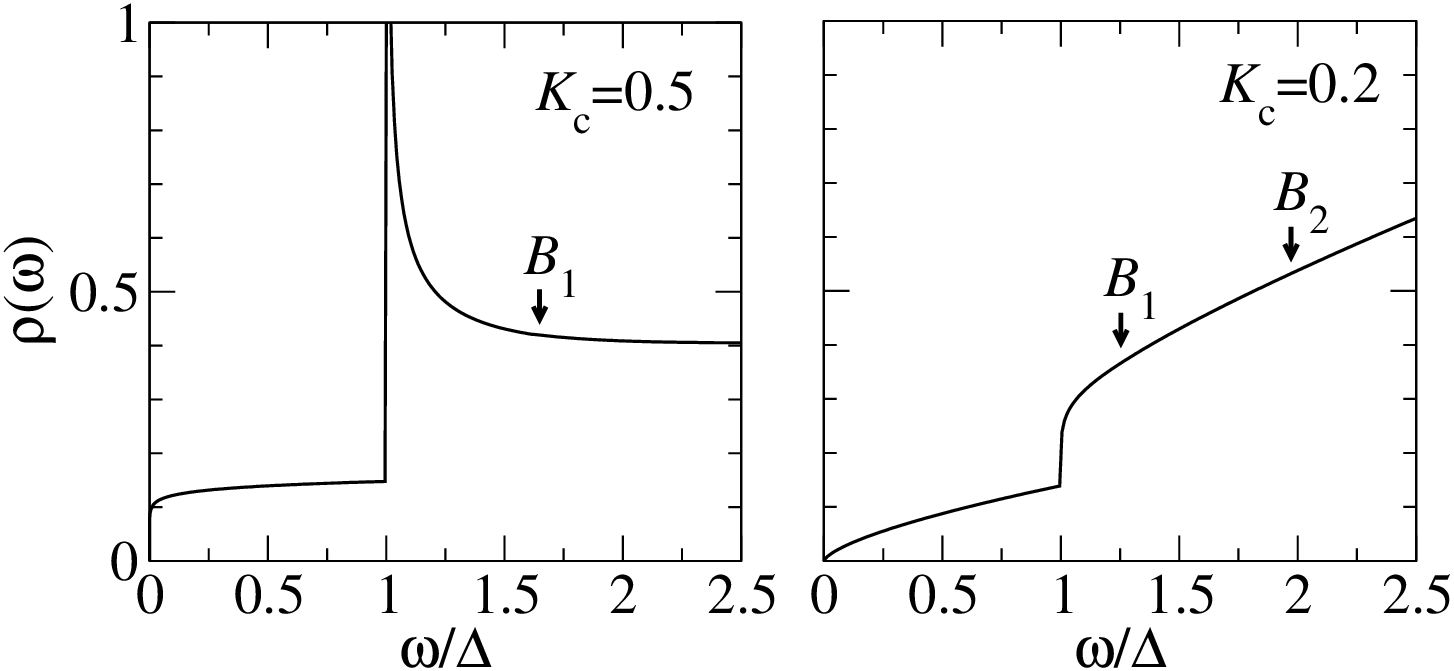}
\caption{LDOS for $K_\text{c}=0.5$ (left panel) and $K_\text{c}=0.2$ (right panel),
$K_\text{s}=1$, and $v_\text{c}=1.2\,v_\text{s}$. We observe a power-law behavior for
$\omega\to 0$ [see Eq.~\eqref{eq:rho1}]. The singularity at $\omega=\Delta$ is only 
present for $K_\text{c}>\sqrt{5}-2$  [see Eq.~\eqref{eq:rho2}]. The arrows indicate 
the energies where single-breather processes set in.}
\label{fig:rho}
\end{figure}

\emph{LDOS.} Scanning tunneling microscopy (STM) and spectroscopy experiments 
measure the tunneling current between the sample and the STM tip as a function of the 
position of the tip and the applied voltage. The tunneling current is directly related to 
the LDOS of the sample, which is thus accessible in STM experiments. The LDOS 
$\rho(\omega)$ is obtained from the spectral function by integrating
over the momentum. The result is spin independent and plotted 
in Fig.~\ref{fig:rho}. We note that the LDOS has recently been calculated\cite{Braunecker-11} 
by using a self-consistent harmonic approximation for the gapped sector, i.e., 
by expanding the cosine in \eqref{eq:plushamiltonian} up to quadratic order. 

The low-energy behavior of the LDOS can be determined from the first term
in \eqref{eq:G1}. We find for $|\omega|<\Delta$
\begin{equation}
\rho(\omega)\propto|\omega|^{\alpha},\;
\alpha=4(b^2+c^2)-1=
\frac{\bigl[1+K_\text{c}(K_\text{s}-2)\bigr]^2}{4K_\text{c}(1+K_\text{c}K_\text{s})},
\label{eq:rho1}
\end{equation}
in agreement with the harmonic approximation.\cite{Braunecker-11} This again 
highlights the existence of spectral weight within the gap in contrast to the MI.\cite{MI}
In addition the LDOS possesses a pronounced feature at $\omega=\Delta$ at which the
gapped sector \eqref{eq:plushamiltonian} starts to contribute. For 
${\omega\to\Delta^+}$ the leading term in \eqref{eq:G2} results in
\begin{equation}
\begin{split}
&\rho(\omega)\propto(\omega-\Delta)^{\beta}+\ldots,\\
&\beta=4d^2-\frac{1}{2}=\frac{1+K_\text{c}^2K_\text{s}(K_\text{s}-2)-2K_\text{c}(1+K_\text{s})}
{4K_\text{c}(1+K_\text{c}K_\text{s})},
\label{eq:rho2}
\end{split}
\end{equation}
where the dots represent the regular background \eqref{eq:rho1}.
We note that except for the non-interacting case, where $\beta=-1/2$,  
the exponent differs from the corresponding result of 
Ref.~\onlinecite{Braunecker-11}, as the harmonic approximation applied there 
does not correctly capture the Lorentz spin of solitons and antisolitons.

The most pronounced feature of the LDOS is found at $\omega=\Delta$, which is identical
to the gap observed in ${A_\downarrow(\omega,k_\text{F}+q)}$. Furthermore, the exponents
$\alpha$ and $\beta$ are fixed by the Luttinger parameters determined from the spectral 
function thus providing a consistency check between the measurements. 
Alternatively, if only the LDOS is accessible but the system is spin-rotational invariant 
($K_\text{s}=1$), there is a fixed relation between $\alpha$ and $\beta$, e.g.,
$\beta$ vanishes for $K_\text{c}<\sqrt{5}-2$. 
In any case the breather contributions to the LDOS are negligible (see Fig.~\ref{fig:rho}).

Due to the translational invariance of the system the 
LDOS shows no signatures of propagating quasiparticles. In order to identify such 
features in STM experiments it is thus neccessary to study the spatial Fourier transform
of the LDOS in systems with boundaries or impurities.\cite{STM}

\begin{figure}[tb]
\centering
\includegraphics[scale=0.35,clip=true]{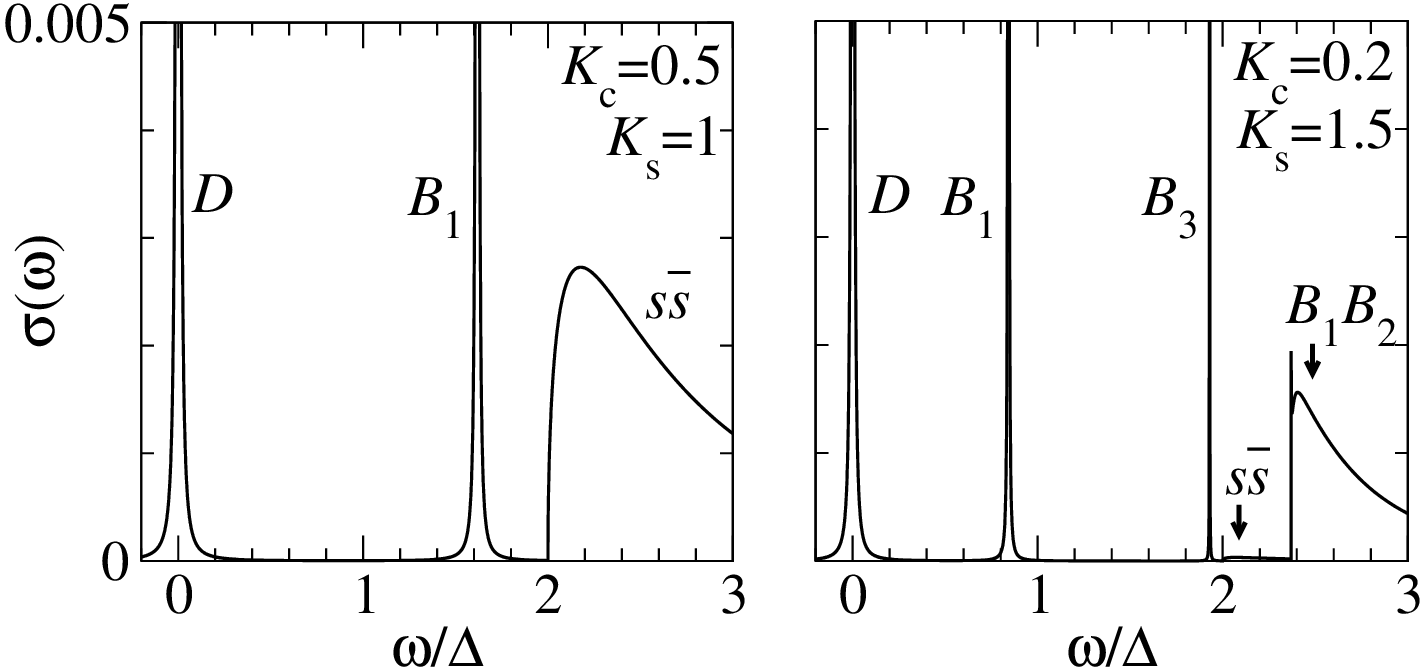}
\caption{Conductivity for $v_\text{c}=1.2\,v_\text{s}$ and $K=1.5$ (left panel) as well as 
$K=0.87$ (right panel). We have convoluted the $\delta$-peaks by a Lorentzian.
In both systems we observe the Drude peak ($D$) at $\omega=0$, 
the $\delta$-peak  from the first breather $B_1$, and the soliton-antisoliton continuum 
($s\bar{s}$). In the right-hand panel we further observe the $\delta$-peak from $B_3$ and 
the $B_1B_2$ continuum, which possesses a singularity at the lower threshold.}
\label{fig:sigma}
\end{figure}
\emph{Conductivity.} Finally let us discuss the optical conductivity\cite{Giamarchi04} 
$\sigma(\omega)$ which is obtained from the current-current correlation function. 
With the current density\cite{Giamarchi04,Braunecker-09}
\begin{equation}
j_e=\frac{e}{\pi}\partial_t\,\Phi_\text{c}
=\frac{e}{\pi\sqrt{K}}\left[K_\text{c}\,\partial_t\,\Phi_+
-\sqrt{\frac{K_\text{c}}{K_\text{s}}}\,\partial_t\,\Phi_-\right]
\end{equation}
it follows that the conductivity is the sum of the conductivities in the gapped 
and gapless sectors, $\sigma(\omega)=\sigma_+(\omega)+\sigma_-(\omega)$. For the 
clean system we consider here the latter is given by\cite{Giamarchi04}  a Drude peak 
$\propto\delta(\omega)$ while the former can be calculated analogously to a 
MI.\cite{Controzzi-01} Thus $\sigma_+(\omega)$ is given by a sum of 
$\delta$-functions originating in single-breather processes and, at higher frequencies,
a many-particle continuum $\sigma_+^\text{cont}(\omega)$.
As the current operator couples only to single-breather states with $n$ odd we find
\begin{equation}
\sigma_+(\omega)=\sum_{j=1}^{(N+1)/2}a_j\,\delta(\omega-\Delta_{2j-1})
+\sigma_+^\text{cont}(\omega),
\end{equation}
where the coefficients $a_j$ are fixed by the single-breather form 
factors\cite{Smirnov92book,EsslerKonik05,Controzzi-01} of the current operator. 

For $4/3\le K<2$ only the breather $B_1$ 
exists. As shown in the left-hand panel in Fig.~\ref{fig:sigma} the conductivity is dominated by 
two $\delta$-peaks at low energies, while at $\omega>2\Delta$ the soliton-antisoliton 
($s\bar{s}$) continuum shows up. In particular the optical gap is twice the gap observed in
the spectral function or the LDOS. The leading corrections are given by the three-particle 
process $s\bar{s}B_1$ which sets in at $\omega=2\Delta+\Delta_1$. 
For $K<4/3$ the second 
breather starts to contribute in the two-breather continuum, but not as a single-particle 
peak as the current operator does not couple to $B_2$. As shown in the right-hand panel 
in Fig.~\ref{fig:sigma}, for $2/3\le K<1$ a third $\delta$-peak originating in $B_3$ exists.
The two-particle continuum is dominated by the two-breather process $B_1B_2$
while the soliton-antisoliton contribution is strongly suppressed as its spectral weight is
transferred to the coherent $B_1$-peak. ($B_1$ is a $s\bar{s}$ bound state.) 
We stress that the $B_1B_2$-continuum possesses a singularity at the threshold 
$\omega\to(\Delta_1+\Delta_2)^+$. The leading corrections are given by the 
three-particle processes $B_1B_1B_1$. We note that the condition $K<1$ requires the
breaking of the spin rotation invariance, i.e., $K_\text{s}>1$. If $K$ is further 
decreased, the conductivity develops more features due to the existence of additional 
breathers.

In contrast to the spectral function and LDOS, the conductivity 
allows an efficient spectroscopy of the individual breather masses. In particular, the 
knowledge of the Luttinger parameters and thus $K$ gives a precise prediction for the
number and masses of detectable individual breather states as well as features in 
the many-particle continuum. This provides a non-trivial consistency check between
measurements of the optical conductivity and the spectral function or LDOS
which has to be satisfied for a definite detection of a SSDW state.

\emph{Conclusions.} In this Rapid Communication 
we have presented a complete characterization 
of the spectral properties of SSDW states, which appear\cite{Braunecker-09} in 
one-dimensional electron systems coupled to nuclear spins or quantum wires in a
magnetic field and with SOI that are commensurate with the Fermi 
momentum.\cite{Braunecker-10} We calculated the spin-resolved spectral function, LDOS, 
and optical conductivity. We identified various collective modes and showed that the
spectrum strongly depends on the interaction strength between the electrons. 
We discussed various consistency checks between the observables and thus 
between measurements of the spectral function, the LDOS, and the optical conductivity 
which should allow an unambiguous detection of a SSDW state in future experiments 
on $^{13}$C single-wall carbon nanotubes and GaAs quantum wires.

I would like to thank Bernd Braunecker, Volker Meden, Pascal Simon, and particularly 
Fabian Essler, for useful discussions. This work was supported by the German Research 
Foundation (DFG) through the Emmy-Noether Program.


\begin{thebibliography}{10}
\bibitem{SLL}
This state was named "spiral Luttinger liquid" in Ref.~\onlinecite{Braunecker-11},
but we stress that it possesses a spectral gap.
\end{thebibliography}
\end{document}